\documentclass[conference]{IEEEtran}
\IEEEoverridecommandlockouts
\usepackage{cite}
\usepackage{amsmath,amssymb,amsfonts}
\usepackage{mathtools}
\usepackage{graphicx}
\usepackage{textcomp}
\usepackage{xcolor}
\usepackage{graphicx}
\usepackage{caption}
\usepackage{subcaption}

\DeclareMathOperator{\diff}{d}
\newcommand{\ddt}{\tfrac{\diff}{\diff \!t}}

\def\BibTeX{{\rm B\kern-.05em{\sc i\kern-.025em b}\kern-.08em
    T\kern-.1667em\lower.7ex\hbox{E}\kern-.125emX}}

\begin{document}
\thispagestyle{plain}

\title{Three-phase grid-forming droop control for unbalanced systems and fault ride through
\thanks{This material is based upon work supported by the Power Systems Engineering Research Center (PSERC) through project S-95.}
}

\author{Prajwal Bhagwat and Dominic Gro\ss{} \thanks{D. Gro\ss{} and P. Bhagwat are with the Department of Electrical 
and Computer Engineering at the University of Wisconsin-Madison, USA; e-mail:dominic.gross@wisc.edu, pbhagwat2@wisc.edu}}

\maketitle

\begin{abstract}
    In this work, we investigate grid-forming (GFM) control for dc/ac voltage source converters (VSC) under unbalanced system conditions and unbalanced faults. To fully leverage the degrees of freedom of VSCs, we introduce the concept of generalized three-phase GFM control that combines individual GFM controls for every phase with a phase balancing feedback. The proposed control allows trading off voltage and power unbalance under unbalanced conditions, enables current limiting for each phase during unbalanced faults, and reduces to positive sequence GFM droop control in balanced systems. High-fidelity simulations are used to illustrate the properties of the control.
\end{abstract}


\section{Introduction}
Power electronics interfacing renewables, storage, and novel transmission technologies are envisioned to be the cornerstone of tomorrow's resilient and sustainable power systems \cite{K_2017}. In theory, state-of-the-art power converter control can replace a wide range of functions and characteristic commonly associated with conventional fuel-based synchronous generators (SGs) (see, e.g., \cite{DSF_2015}). However, standard controls available in literature often neglect crucial aspects such as unbalanced system conditions, converter current limits, and fault ride through. Ultimately, this approach jeopardizes system reliability and resilience and represents a significant barrier for large-scale integration of renewables and power electronics.

Control strategies for grid-connected power converters can be broadly categorized into (i) grid-following strategies that may provide grid-supporting services but require other devices to form a stable ac voltage waveform that grid-following (GFL) converters can lock on to, and (ii) grid-forming (GFM) control strategies that impose a well-defined and stable ac voltage waveform at their point of connection \cite{CDA_1993,DSF_2015,GCB_2019} and are envisioned to replace SGs as the foundation of future sustainable power systems \cite{M_2019,CTG_2020}. {Key advantages of GFM control include its fast response to contingencies \cite{CTG_2020} and potential for fast fault current injection. In contrast, SGs provide a fault response that is constrained by the electromechanical limits of SGs \cite{K_2017,NCO+2014}. Due} to their comparatively low current limits, power converters cannot emulate the fault response of synchronous machines \cite{QGC_2020}. Consequently, the fault-response of a converter-dominated power system may vastly differ from the response expected by today's {system protection \cite{JYN_2019}. Several strategies for GFL converters} leverage their current source characteristics during faults \cite{JYN_2018}. However, the (self-synchronizing) voltage source characteristics of GFM converters are not amenable to this approach.

Specifically, GFM controls provide a balanced voltage reference that is tracked by underlying cascaded voltage and current controllers. Strategies for current limiting in GFM voltage source converters (VSCs) include limiting the reference of the inner current controller \cite{QGC_2020}, threshold virtual impedance \cite{PD_2015,QGC_2020}, and projected droop control \cite{GD_2019}. However, all the aforementioned works assume a balanced system and balanced faults. At the same time, the vast majority of faults in high voltage systems are unbalanced \cite{IEEE_1992}. 

Moreover, distribution systems typically exhibit significant unbalance that is not accounted for in the vast majority of GFM controls. The few works that consider GFM control under unbalanced conditions \cite{ASD_2021,ARY_2021} and faults \cite{ARY_2021} typically leverage symmetrical components, e.g., implementing separate GFM controls for positive and negative sequence. This approach is motivated by prevailing analysis methods for unbalanced systems using symmetrical components and can effectively control the VSC terminal voltage under unbalanced conditions. However, the relationship between the converter phase currents and symmetrical components is highly nonlinear \cite{ARY_2021} and limiting the phase currents through control of symmetrical components results in challenging control design and analysis problems. {Instead, we propose explicitly controlling the phase voltages and currents in their original coordinates. This results in a less complex controller that can readily use standard current limiting algorithms and result in less complex control design problems and analysis problems.}

{The} contribution of this work is a generalized three-phase GFM droop control for two-level dc/ac VSCs that leverages the capabilities of VSCs to control individual phase voltages and currents during unbalanced conditions and faults. To this end, single-phase GFM controls for every phase are combined with a phase balancing feedback. Using this approach, the (outer) GFM control can provide an unbalanced voltage reference under unbalanced conditions or unbalanced faults. Notably, the proposed control allows trading off voltage unbalance and power unbalance and reduces to standard balanced GFM droop control in balanced systems. The GFM voltage references are tracked by individual inner current and voltage controllers for every phase with phase current limiting. As a result, the proposed generalized GFM control maintains control over the VSC terminal voltage under unbalanced conditions and phase currents under unbalanced faults. {Finally, these features} are illustrated using a high-fidelity EMT simulation.

\section{Converter model and control objectives}
This work considers control of a grid-connected two-level three-phase dc/ac VSC shown in Fig.~\ref{fig:stdgfm} and Fig.~\ref{fig:3phgfm}. For brevity of the presentation, we assume that the midpoint of the dc bus is grounded and that the three-phase ac voltage $v_{\text{sw}}\in\mathbb{R}^3$ modulated by the VSC is a control input (i.e., we consider an average model and neglect the dc bus dynamics). Moreover, we use $i\in\mathbb{R}^3$, $v \in \mathbb{R}^3$, and $i_o \in \mathbb{R}^3$ to denote the filter current, terminal voltage, and output current, respectively. The key control objectives of GFM control are to
\begin{enumerate}
 \item impose a stable ac voltage waveform at its terminal,
 \item asymptotically track a steady-state dispatch signal by a higher-level controller under nominal system conditions,
 \item autonomously respond to contingencies as well as variations in load and generation to stabilize the system, and
 \item operate within the VSC limits (e.g., current limit).
\end{enumerate}
The first three objectives are in line with envisioning GFM VSCs as replacement for SGs. However, the low overcurrent capability of VSCs relative to SGs preclude emulating the response of a SG during severe contingencies such as faults. This fact poses a challenge for prevailing protection paradigms and analysis methods. {While resolving these  challenges is beyond the scope of this manuscript, we propose a generalized GFM control that fully leverages the degrees of freedom of VSCs under unbalanced conditions and faults.}

\subsection{Standard (positive sequence) GFM control}
\begin{figure}[h!!]
    \centering
    \includegraphics[width=0.95\columnwidth]{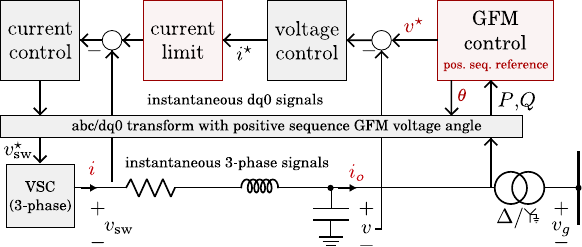}
    \caption{{Standard dual-loop GFM control with inner controls tracking a positive sequence voltage reference provided by an outer GFM control (e.g., droop control \cite{CDA_1993}, VSM \cite{DSF_2015}).}\label{fig:stdgfm}}
\end{figure}

To motivate our contribution, we briefly discuss the standard dual-loop GFM control architecture shown in Fig.~\ref{fig:stdgfm} in which the GFM control provides a positive sequence voltage reference with angle $\theta^{\text{gfm}} \in \mathbb{S}$ and magnitude $V^{\text{gfm}} \in \mathbb{R}$ that is tracked by inner proportional-integral (PI) current and voltage controls in a $dq$-frame attached to the angle $\theta^{\text{gfm}}$. {For brevity of the presentation,} we consider a standard reference limiter for overcurrent protection (e.g., during faults) implemented in the same $dq$-frame. {Notably, the standard dual-loop GFM control architecture was designed for balanced systems. Because of this, the standard inner loops can typically not track the balanced GFM voltage reference during unbalanced conditions and faults and may fail to limit the VSC phase currents.}

\section{Generalized three-phase GFM control}
\subsection{Control architecture}\label{subsec:architecture}
\begin{figure}[h!!]
    \includegraphics[width=0.95\columnwidth]{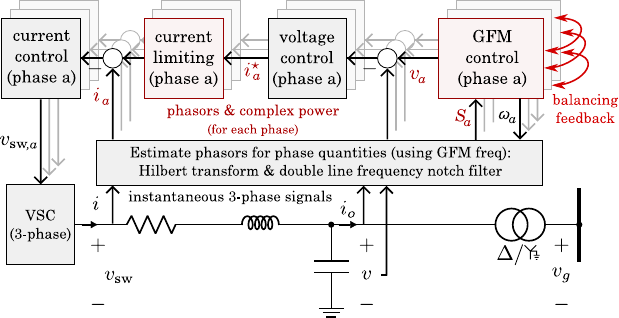}
    \caption{{Generalized three-phase GFM control with inner control for every phase tracking a voltage reference provided by the outer GFM control \eqref{eq:gendroop} with phase balancing feedback.}\label{fig:3phgfm}}
\end{figure}

Figure~\ref{fig:3phgfm} shows the generalized three-phase GFM control architecture proposed in this paper that combines individual inner current and voltage controllers for every phase, individual GFM controls for every phase, and a phase balancing feedback. To enable control of individual phase currents and voltages, a reference frequency $\omega^{\text{gfm}}_p \in \mathbb{R}$ and angle $\theta^{\text{gfm}}_p \in \mathbb{S}$ for every phase $p \in \mathcal{P}$ with $\mathcal P \coloneqq \{a,b,c\}$ provided by the outer GFM control is used to estimate phasors for the phase voltages and currents. Using these estimates, the voltage controller tracks a voltage reference for every phase. The resulting current reference for every phase is limited individually and tracked by individual current controls. {Analogously to positive sequence GFM control, the gains of inner and outer loops need to be coordinated and chosen relative to the network circuit dynamics to ensure performance and stability \cite{QGC_2018,GCB_2019}.} The (outer) GFM control combines GFM droop control for every phase with a phase balancing feedback that (i) ensures balanced references if the system is balanced, and (ii) allows trading off voltage and power unbalances  under unbalanced conditions and faults.

\subsection{Phasor extraction and phase powers}\label{subsec:phsext}
To control the voltages $v_p \in \mathbb{R}$ and currents $i_p \in \mathbb{R}$ for each phase $p \in \mathcal P$, we require their quadrature components and magnitude. To this end, let $\omega_p \in \mathbb{R}$ denote the GFM reference voltage frequency for phase $p \in \mathcal P$. Then, for all $p \in \mathcal P$ and any ac signal $x(t)$, we estimate the quadrature component $x^\perp_p(t) \in \mathbb{R}$ of $x_p(t) \in \mathbb{R}$ as
\begin{align}
     x^\perp_p(t) \coloneqq x_p(t-\tfrac{1}{4}\tfrac{1}{\omega_p(t)}). \label{eq:ht:perp}
\end{align}
In other words, under the assumption that $x_p(t)$ is a sinusoidal signal with slowly changing frequency $\omega_p(t)$, the time shifting in \eqref{eq:ht:perp} referred to as Hilbert transform in the power electronics literature approximates a $90^\circ$ phase shift. Notably, the Hilbert transform can be interpreted as representing $\alpha\beta$ components of an ac signal (i.e., $(x_{p,\alpha},x_{p,\beta})=(x_p,x^\perp_p)$). Let $R(\cdot)$ denote the 2D rotation matrix. Then, any ac signal $x_p(t)$ can be represented in a $dq$ frame with reference angle $\theta_p(t) \in \mathbb{S}$ as
\begin{align}
    x_{p,dq}(t)\coloneqq\begin{bmatrix} x_{p,d}(t) \\ x_{p,q}(t) \end{bmatrix} \coloneqq R(\theta_p(t)) \begin{bmatrix} x_p(t) \\ x_p^\perp(t) \end{bmatrix}.
\end{align}
Moreover, for any ac signal $x_p(t)$ we can construct the phasor
\begin{align}
    \mathbf{x}_p(t) \coloneqq x_{p,d}(t) + j x_{p,q}(t). 
\end{align}
Finally, for all phases $p \in \mathcal P$, the average active power $P_p$ and reactive power $Q_p$ over one cycle can be computed using the filter current $\mathbf{i}_p(t)$, voltage phasor $\mathbf{v}_p(t)$, and
\begin{subequations}\label{eq:ht_pq}
    \begin{align}
        P_p(t) \coloneqq \tfrac{1}{2} \operatorname{Re}(\mathbf{v}_p \overline{\mathbf{i}_p}) = \tfrac{1}{2} \left(v_p(t) i_p(t) + v_p^\perp(t) i_p^\perp(t)\right), \label{eq:ht_pq:p}\\
        Q_p(t) \coloneqq \tfrac{1}{2} \operatorname{Im}(\mathbf{v}_p \overline{\mathbf{i}_p}) = \tfrac{1}{2} \left(v_p^\perp(t) i_p(t) - v_p(t) i_p^\perp(t)\right). \label{eq:ht_pq:q}
    \end{align}
\end{subequations}
{Because of controller sampling rate limits the time-shift \eqref{eq:ht:perp} cannot be implemented to arbitrary accuracy. This results in a double line frequency component in the measurements of $P_p$ and $Q_p$ that is removed using a notch filter with center frequency $2 \omega_p$. In the remainder, we omit the time variable $t$.}

\subsection{Generalized three-phase droop control}\label{subsec:out_cnt}
The main contribution of this work is a generalized GFM droop control that combines single-phase droop control for every phase $p\in\mathcal P$ with a phase balancing feedback. Considering $(\theta^{\text{bal}}_a,\theta^{\text{bal}}_b,\theta^{\text{bal}}_c)\coloneqq (0,\tfrac{2}{3}\pi,-\tfrac{2}{3}\pi)$ and voltage setpoints $V^\star_p \in \mathbb{R}_{\geq 0}$, the GFM voltage angle and magnitude references $\theta^{\text{gfm}}_p = \delta^{\text{gfm}}_p + \theta^{\text{bal}}_p$ and $V^{\text{gfm}}_p = V^{\text{gfm}}_{\delta,p} + V^\star_p$ are determined by the outer GFM control in per unit
\begin{subequations}\label{eq:gendroop}
    \begin{align}
    \ddt \delta^{\text{gfm}}_p \!&=\! \omega_0-\!\sum_{l\in \mathcal P\setminus p}\!\! k_s(\delta^{\text{gfm}}_p\!-\!\delta^{\text{gfm}}_l)\!+\!m_P(P_p^{\star}\!-\!P_p),
\label{eq:3dr_p}\\
\!\!\tau \!\ddt \!V^{\text{gfm}}_{\delta,p} \!&=\! -V^{\text{gfm}}_{\delta,p}-\!\!\!\sum_{l\in \mathcal P\setminus p}\!\!\! k_s(V^{\text{gfm}}_{\delta,p}\!-\!V^{\text{gfm}}_{\delta,l})\!+\!m_Q(Q_p^{\star}\!-\!Q_p).\!\! \label{eq:3dr_q}
    \end{align}
\end{subequations}
Here, $P^\star_p \in \mathbb{R}$ and $Q^\star_p \in \mathbb{R}$  are the active and reactive power setpoints and $\omega_0 \in \mathbb{R}_{>0}$ is the nominal frequency. Moreover, $\omega^{\text{gfm}}_p = \omega_0 + \ddt \delta^{\text{gfm}}_p \in \mathbb{R}$ is the GFM reference frequency for each phase $p\in\mathcal P$, $m_P \in \mathbb{R}_{> 0}$ is the $P-f$ droop coefficient, $m_Q \in \mathbb{R}_{>0}$ is the $Q-V$ droop coefficient, $\tau \in \mathbb{R}_{>0}$ a lowpass filter time constant, and $k_s \in \mathbb{R}_{\geq 0}$ is the phase balancing gain. 

\begin{figure}[t]
    \begin{subfigure}[b]{1\columnwidth}
        \centering
        \includegraphics[width=0.85\columnwidth]{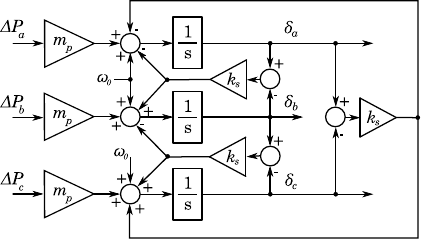}
        \caption{Generalized three-phase $P-f$ droop.}
        \label{fig:pfdroop}
    \end{subfigure}
\vspace{0.2em}

    \begin{subfigure}[b]{1\columnwidth}
        \centering
        \includegraphics[width=0.85\columnwidth]{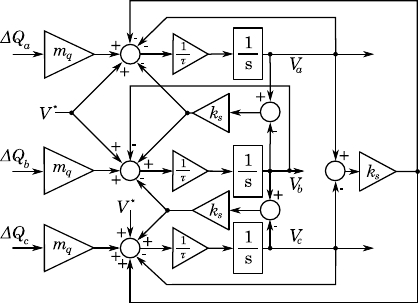}
        \caption{Generalized three-phase $Q-V$ droop.}
        \label{fig:qvdroop}
    \end{subfigure}
    \caption{Generalized three-phase GFM droop control.}\label{fig:gendroop}
\end{figure}

We emphasize that \eqref{eq:gendroop} reduces to three individual single-phase droop controls if $k_s=0$, i.e., the GFM voltage reference for each phase $p \in \mathcal P$ only depend on the power measurements of phase $p \in \mathcal{P}$. In contrast, if $k_s >0$, the phase balancing feedback trades off voltage phase unbalance and deviations from the power setpoints for each phase. Finally, for large $k_s$ the phase voltage balancing is stiff and the response of generalized three-phase droop control converges to the response of standard GFM droop control  as $k_s \to \infty$. A block diagram of the GFM control \eqref{eq:gendroop} is shown in Fig.~\ref{fig:gendroop}.
The GFM voltage references are tracked by inner current and voltage controls described in the next section.

\subsection{Dual-loop current/voltage control and current limiting}\label{subsec:inn_cnt}
We extend the standard dual-loop proportional integral (PI) current and voltage control with reference current limiting \cite{QGC_2020} to track the GFM voltage references provided by \eqref{eq:gendroop} for each phase (see Fig.~\ref{fig:3phgfm}). Specifically, the voltage controller
\begin{align}\label{eq:voltloop}
    i^{\text{ref}}_{p,dq} = i_{o,p,dq} + Y_f v_{p,dq} + G_{\text{PI}}(s)(v^{\text{gfm}}_{p,dq}-v_{p,dq})
\end{align}
with PI controller $G_{\text{PI}}(s)$, filter admittance matrix $Y_f \in \mathbb{R}^2$, and voltage reference $v^{\text{gfm}}_{p,dq}=(V^{\text{gfm}}_p,0)$ computes the filter current reference $i^{\text{ref}}_{p,dq}$ for all $p \in \mathcal{P}$. Next, using the maximum phase current magnitude $I_{\max} \in \mathbb{R}_{>0}$, the reference $\mathbf{i}^{\text{lim}}_p$ for to the inner current loop of phase $p \in \mathcal{P}$ is limited by
\begin{align}\label{eq:curr_lim}
   \mathbf{i}^{\text{lim}}_p \coloneqq
    \begin{cases}
        \mathbf{i}^{\text{ref}}_p & \quad \textrm{if } \lVert \mathrm{i}^{\text{ref}}_p \rVert \leq I_{\max} \\
        i_{\max} \angle \mathbf{i}^{\text{ref}}_p & \quad \textrm{if }  \lVert \mathrm{i}^{\text{ref}}_p \rVert > I_{\max}
    \end{cases}.
\end{align}
Notably, unlike some commercial solutions, the current limiter \eqref{eq:curr_lim} does not clip the current waveform, but adjusts the magnitude of the sinusoidal reference current for every phase to avoid introducing harmonics into the system. Finally, for each $p \in \mathcal{P}$ we use the inner current controller 
\begin{align}\label{eq:currloop}
    v_{\text{sw},p,dq} = v_{p,dq} + Z_f i_{p,dq} + G_{\text{PI}}(s)(i^{\text{lim}}_{p,dq}-i_{p,dq}),
\end{align}
where $Z_f \in \mathbb{R}^2$ denotes the filter impedance matrix and $v_{\text{sw},p} \in \mathbb{R}$ is the phase voltage modulated by the VSC.

\subsection{Discussion}
The generalized three-phase droop control proposed in this section has {three} key features. First, the balancing gain $k_s$ adjusts the trade-off between phase voltage unbalance and power unbalance at the converter ac terminal. For example, $k_s$ allows to adjust the contribution of a dc/ac VSC to mitigating voltage unbalances in a distribution system. Second, even for large values of $k_s$, a key feature of generalized three-phase droop control is that it (i) tracks voltage references for every phase, and (ii) can control and limit the phase currents individually. {Third, the control explicitly addresses sub-cycle overcurrent by continuously estimating and controlling phase current phasors and limiting their magnitude. Finally, from a theoretical point of view, the proposed control reduces to standard GFM control if the system is balanced. Simulations performed to compare (positive sequence) performance and stability to standard GFM control only identified significant differences when using negligible phase-balancing gains $k_s$.}

\section{Case study}
\subsection{System description}
\begin{figure}[h]
    \centering
    \includegraphics[width=0.99\columnwidth]{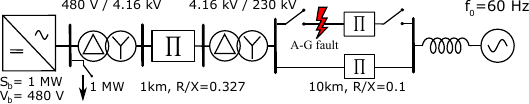}
    \caption{Test system with a low-voltage VSC connected to a weak ac system through a distribution line, double circuit transmission line, and a step up transformer.\label{fig:invconn}}
\end{figure}

{To illustrate the results, we use a high-fidelity EMT simulation of a two-level dc/ac VSC connected to an infinite bus through a $1$~km medium voltage line, a $10$~km double circuit high voltage transmission line, and step up transformers. The system parameters are shown in Fig.~\ref{fig:invconn} and the control ($m_P\!=\!m_Q\!=\!5\%$) is implemented at a sampling rate of $10~\mathrm{kHz}$.}

\subsection{Unbalanced load}\label{unb_load}
\begin{figure}[h]
    \centering
    \includegraphics[width=0.99\columnwidth]{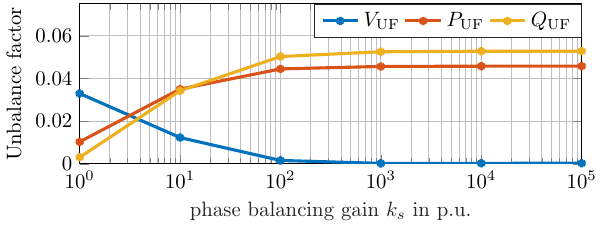}
    \caption{Steady-state unbalance factors for an unbalanced load at the VSC terminal and various phase balancing gains $k_s$. \label{fig:ks_resp}}
\end{figure}

To illustrate the role of the phase balancing gain $k_s$, let $V^{+}$ and $V^-$ denote  magnitude of the positive and negative sequence voltage and let $\bar{P} \coloneqq \tfrac{1}{3}\sum_{p\in\mathcal{P}} P_p$ and $\bar{Q}\coloneqq \tfrac{1}{3}\sum_{p\in\mathcal{P}} Q_p$ denote the average phase power of the VSC. The standard voltage unbalance factor is given by $V_{\text{UF}} \coloneqq V^-/V^+$. Moreover, for $P_p$ and $Q_p$ in p.u., we define the power unbalance factors $P_{\text{UF}} \coloneqq \max_{p \in \mathcal{P}} \| P_p - \bar{P} \|$, and $Q_{\text{UF}} \coloneqq \max_{p \in \mathcal{P}} \| Q_p - \bar{P}\|$ that resemble standard current unbalance factors for electric machines. 
Figure~\ref{fig:ks_resp} shows the steady-state {voltage and power} unbalance factors {at} the VSC terminal {for} an unbalanced delta connected constant impedance load\footnote{Relative to the load between phase $a$ and $b$, the load between phase $b$ and $c$ is 20\% lower and the load between phase $a$ and $c$ is 20\% higher} at the VSC bus (see Fig.~\ref{fig:invconn}). The results show the expected trade off between voltage unbalance and power unbalance, i.e., increasing the phase balancing gain $k_s$ reduces voltage unbalance $V_{\text{UF}}$ but increases power unbalance $P_{\text{UF}}$ and $Q_{\text{UF}}$.
\subsection{Single line-to-ground fault}
\begin{figure}[h]
    \centering
    \includegraphics[width=0.95\columnwidth]{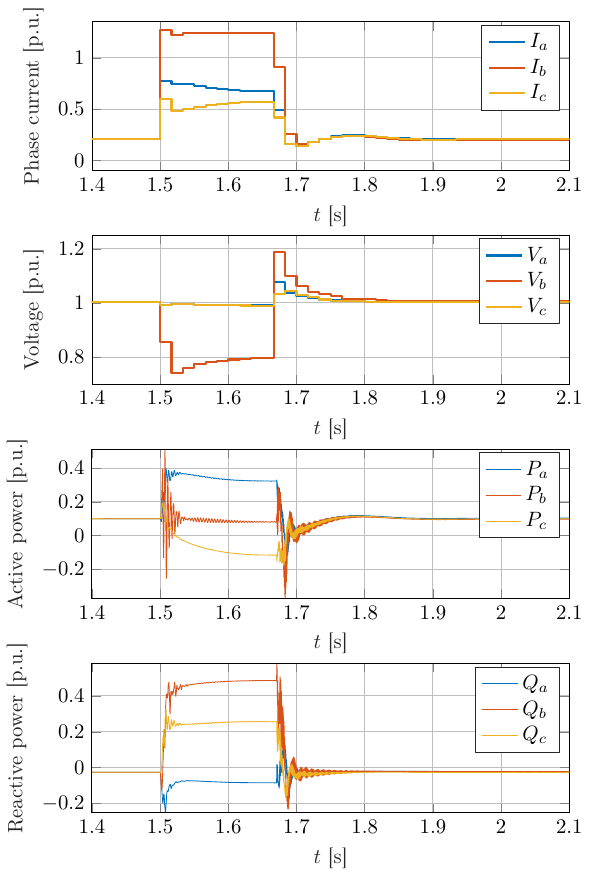}
    \caption{Response of generalized droop control to a phase $a$ to ground fault  on a transmission line at $t=1.5~\mathrm{s}$. The fault is cleared after ten cycles by disconnecting the faulted line.\label{fig:invresp}}
\end{figure}

To illustrate the impact of separate voltage/current control and current limiting for every phase on unbalanced fault ride through, we disconnect the load at the VSC terminal and consider a zero impedance line-to-ground fault for phase $a$ of a transmission line (see Fig.~\ref{fig:invconn}). {Figure~\ref{fig:invresp} shows the resulting VSC terminal voltage magnitude $V_p=\|\mathbf{v}_p\|$, VSC phase current magnitude $I_p=\|\mathbf{i}_p\|$, active power $P_p$, and reactive power $Q_p$ for every phase $p \in \mathcal P$. Because the magnitudes of phase currents and voltages are not well-defined within a cycle, the maximum magnitude over one cycle is shown.} We use $k_s=10^5~\mathrm{p.u.}$, i.e., the outer GFM control is configured to resemble standard droop control, and the fault is applied at $t=1.5~\mathrm{s}$ and cleared after ten cycles by disconnecting the faulted line. Notably, due to the transformer connection, the fault applied to phase $a$ of the transmission line, is effectively mapped to phase $b$ at the VSC terminal.
It can be seen that the current reference limiter \eqref{eq:curr_lim} and PI current control \eqref{eq:currloop} successfully limits the current magnitude to $I_{\max}=1.2~\mathrm{p.u.}$ {within each cycle. Notably, by controlling the current phasor for every phase, the proposed control explicitly handles sub-cycle overcurrent.} Once the fault is cleared a significant resynchronization transient is observed that can be attributed to controller windup in the  voltage loop \eqref{eq:voltloop} and GFM angle dynamics \eqref{eq:3dr_p}. {Combining the proposed per-phase control with advanced current limiting schemes such as threshold virtual impedance \cite{QGC_2020}} is seen as interesting topic for future work.  

\begin{figure}[h]
    \centering
    \includegraphics[width=0.95\columnwidth]{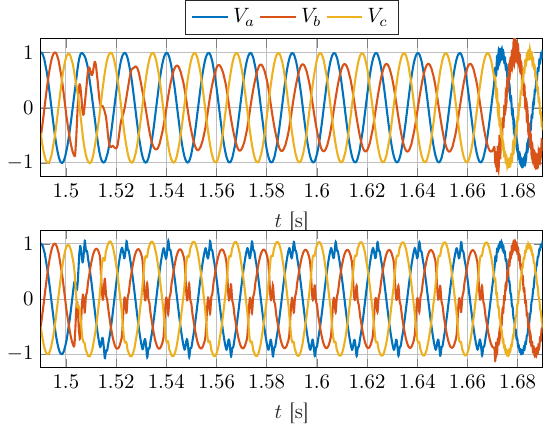}
    \caption{VSC voltages during a phase $a$ to ground fault using generalized three-phase droop control with dual-loop current/voltage control for every phase (top) and standard droop control and  dual-loop current/voltage control (bottom). \label{fig:vresp}}
\end{figure}

To further illustrate the positive impact of the generalized three-phase GFM control, the  simulation study has been repeated with standard droop control using the dual-loop inner control structure. In this case, the current and voltage waveforms are severely distorted because the VSC aims to impose balanced phase currents and a balanced voltage at its terminal. However, under its current limits, the VSC cannot maintain a balanced voltage at the terminal during the unbalanced fault. Terminal voltage waveforms for both controls are shown in Fig.~\ref{fig:vresp}. It can be seen that the proposed control successfully imposes a sinusoidal ac voltage waveform at the terminals of the VSC that is unbalanced to limit the current. In contrast, standard droop control results in a highly distorted voltage.

\section{Conclusion}
{This paper considered grid-forming control for three-phase dc/ac voltage source converters under unbalanced conditions and faults.} In this context, we proposed a novel generalized three-phase droop control that  combines individual GFM controls for every phase with a phase balancing feedback. This generalized GFM control fully leverages the degrees of freedom of the VSC and allows trading off voltage and power unbalance under unbalanced conditions and reduces to standard droop control in balanced systems. Moreover, the voltage references generated by the outer GFM controls are tracked by inner dual-loop current and voltage controls for each phase that enable direct limiting of phase currents without clipping the current waveforms. In contrast to standard GFM controls, this approach maintains full control of the VSC terminal voltage during unbalanced faults and enables effective limiting of phase currents during unbalanced faults. A high-fidelity {case study} is used to demonstrate the effectiveness of the proposed control under unbalanced low voltage loads and unbalanced faults{. Topics for future work include (i) in-depth comparisons to standard GFM control using standard performance and stability metrics, (ii) studying interactions of GFM control with legacy protection systems, and (iii) extending advanced current limiting schemes to enable their per-phase application in generalized three-phase GFM control.}

\bibliographystyle{IEEEtran}
\bibliography{IEEEabrv,bibly}

\end{document}